\documentclass[aps,prl,twocolumn,10pt,showpacs,superscriptaddress,amsmath,amssymb,nobibnotes]{revtex4-1}

\usepackage{graphicx}
\usepackage{amsfonts}    
\usepackage{graphicx}   
\usepackage{verbatim}   
\usepackage{color}      
\usepackage{subfigure}  
\usepackage{hyperref}   
\usepackage{natbib}

\begin{document}

\title{Experimental creation of superposition of unknown photonic quantum states}
\author{Xiao-Min Hu}\email{These authors contributed equally to this work}
\author{Meng-Jun Hu}\email{These authors contributed equally to this work}
\author{Jiang-Shan Chen}
\author{Bi-Heng Liu}\email{bhliu@ustc.edu.cn}
\author{Yun-Feng Huang}
\author{Chuan-Feng Li}
\author{Guang-Can Guo}
\author{Yong-Sheng Zhang}\email{yshzhang@ustc.edu.cn}
\affiliation{Laboratory of Quantum Information, University of Science and Technology of China, Hefei, 230026, China}
\affiliation{Synergetic Innovation Center of Quantum Information and Quantum Physics, University of Science and Technology of China, Hefei, 230026, China}

\date{\today}

\begin{abstract}
As one of the most intriguing intrinsic properties of quantum world, quantum superposition provokes great interests in its own generation. Oszmaniec {\it et al.} [Phys. Rev. Lett. {\bf 116}, 110403 (2016)] have proven that though a universal quantum machine that creates superposition of arbitrary two unknown states is physically impossible, a probabilistic protocol exists in the case of two input states have nonzero overlaps with the referential state. Here we report a heralded quantum machine realizing superposition of arbitrary two unknown photonic qubits as long as they have nonzero overlaps with the horizontal polarization state $|H\rangle$. A total of 11 different qubit pairs are chosen to test this protocol by comparing the reconstructed output state with theoretical expected superposition of input states. We obtain the average fidelity as high as 0.99, which shows the excellent reliability of our realization. This realization not only deepens our understanding of quantum superposition but also has significant applications in quantum information and quantum computation, e.g., generating non-classical states in the context of quantum optics and realizing  information compression by coherent superposition of results of independent runs of subroutines in a quantum computation.
\end{abstract}

\maketitle
{\it Introduction.}
Quantum superposition, which makes quantum world totally different from classical world, is at the heart of quantum theory \cite{Dirac}. The superposition of states leads to inevitable uncertainty in the measurement outcomes, which is the fundamental feature of quantum world. Numerous nonclassical properties of quantum system such as quantum coherence \cite{coh1, coh2} and quantum entanglement \cite{Hor}, which are foundations of quantum communication and computation \cite{Gisin, Wineland}, also essentially stem from quantum superposition. The importance of quantum superposition is exhibited by not only its fundamental role in quantum theory but also significant applications in quantum information and quantum computation \cite{Chuang}.

As a fascinating consequence of linearity of quantum theory, quantum superposition raises great interest in its own generation, i.e., whether or not there exists a universal quantum machine that produces superposition of arbitrary two unknown input quantum states \cite{Solano, Mik}. Unfortunately, similar to other no-go theorems \cite{go1, go2, go3, go4}, such universal protocol has been shown forbidden by quantum theory \cite{Solano, Mik}.  However, it is similar to probabilistic quantum cloning machine for linear-independent set of states \cite{Duan1, Duan2} that probabilistically creating superposition of arbitrary two states is feasible given that both states have nonzero overlaps with some referential states \cite{Mik}.

In this Letter, we experimentally demonstrate a heralded probabilistic quantum machine realizing superposition of arbitrary two photonic qubits based on the protocol in Ref. \cite{Mik}. The referential state in our experiment is conveniently chosen to be horizontal polarization state $|H\rangle$ of photon, so that arbitrary two qubits can be superposed except for that in vertical polarization state $|V\rangle$. We test the superposition machine that is named quantum adder hereafter by inputting some representative qubit states and performing tomography of the output states of superposition. We use fidelity to denote the similarity between the observed and the expected superposition states. The average fidelity is obtained as high as 0.99, which shows the high reliability of our quantum adder machine.

\begin{figure}[bp]
\centering
\includegraphics[scale=0.4]{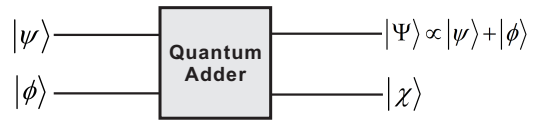}
\caption{Schematic of the quantum adder. A quantum adder outputs a superposed state $|\Psi\rangle\propto|\psi\rangle+|\phi\rangle$ for two arbitrary input states $|\psi\rangle$ and $|\phi\rangle$ with some probability given that $\langle\chi|\psi\rangle\neq 0,\langle\chi|\phi\rangle\neq 0$.}
\end{figure}


\begin{figure}[tbp]
\centering
\includegraphics[scale=0.48]{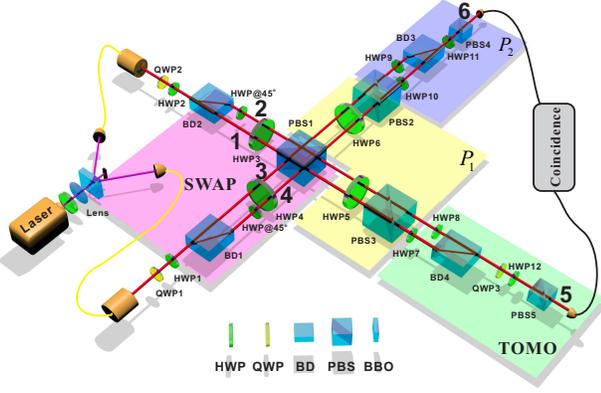}
\caption{(Color online). Experimental setup for realizing quantum adder. A pair of photons in state $|H\rangle\otimes|H\rangle$ is generated via spontaneous parametric down-conversion (SPDC) process by pumping a type-I cut $\beta$-BBO crystal with ultraviolet (UV) laser at $404$ nm. By using half wave plates (HWP1, HWP2) and quarter wave plates (QWP1, QWP2) placed before beam displacers (BD1, BD2), arbitrary two photons' state $(a|H\rangle+b|V\rangle)\otimes(c|H\rangle+d|V\rangle)$ can be produced. BD1, BD2, HWP3, HWP4 (HWPs are fixed at $22.5^{\circ}$) and a polarizing beam splitter (PBS1) together with the detection of coincident events realize the control-SWAP operation. BD1, BD2, HWP3 and HWP4 are used to transform polarization qubits into path qubits. PBS1 placed at the center of the setup combined with selection of coincidence realizes the swap operation between different path qubits with control state $|H\rangle_{1}|H\rangle_{2}+|V\rangle_{1}|V\rangle_{2}$. Path qubits are transformed back into polarization qubits through HWP7, HWP8, HWP9, HWP10, BD3 and BD4, in which HWP7 and HWP9 are fixed at $45^{\circ}$ while HWP8 and HWP10 are fixed at $0^{\circ}$ for optical path compensation. $P_{1}$ section stands for the projection operation of control qubit by setting HWP5 and HWP6 at $22.5^{\circ}$, and only allowing photons in horizontal polarization state $|H\rangle$ to pass through PBS2 and PBS3. HWP11 and PBS4 in section $P_{2}$ realize the projection operation of photon 2. QWP3 and HWP12 combined with PBS5 are used for state tomography of the output state. The coincident events herald the success of superposing two input qubits.}
\end{figure}

{\it Realization of quantum superposition.}
We begin by a brief review of the basic idea of realization of quantum superposition \cite{Mik}. Suppose that there are two photons in unknown normalized states $|\psi\rangle_{1}=a|H\rangle+b|V\rangle$ and $|\phi\rangle_{2}=c|H\rangle+d|V\rangle$ with subscripts to distinguish the two photons. The aim of the quantum adder is to transform the two photons' states $|\psi\rangle$ and $|\phi\rangle$ into a single superposition state $|\Psi\rangle\varpropto |\psi\rangle+|\phi\rangle$ with the help of ancilla (Fig. 1). This transformation can not be realized only by unitary operation. The reason is that if this is true, its inverse process implies the possibility of cloning an arbitrary state, which obviously violates the no-cloning theorem \cite{Solano}.

To realize quantum superposition, an entangled operation must act on the two input states. This operation is first realized by the control-SWAP between states $|\psi\rangle_{1}$ and $|\phi\rangle_{2}$ with the help of the normalized control qubit $\alpha|0\rangle+\beta|1\rangle$,
\begin{equation}
U_{swap}|\psi\rangle_{1}|\phi\rangle_{2}(\alpha|0\rangle+\beta|1\rangle)=\alpha|\psi\rangle_{1}|\phi\rangle_{2}|0\rangle+\beta|\phi\rangle_{1}|\psi\rangle_{2}|1\rangle,
\end{equation}
and then the control qubit is projected onto some fixed state say $|+\rangle=(|0\rangle+|1\rangle)/\sqrt{2}$. The composite state of the two photons becomes (unnormalized)
\begin{equation}
|\Phi\rangle_{12}=\alpha|\psi\rangle_{1}|\phi\rangle_{2}+\beta|\phi\rangle_{1}|\psi\rangle_{2}.
\end{equation}

The superposition state now can be obtained by projecting the state of photon 2 onto the referential state $|\chi\rangle$ that $|\langle\chi|\psi\rangle_{2}|\neq 0$ and $|\langle\chi|\phi\rangle_{2}|\neq 0$ are definite. Here we set $|\chi\rangle=|H\rangle$ and we can obtain the superposition state
\begin{equation}
|\Psi\rangle_{1}=\dfrac{1}{N}(\alpha c|\psi\rangle+\beta a|\phi\rangle),
\end{equation}
where $a=\langle H|\psi\rangle_{2}\neq 0, c=\langle H|\phi\rangle_{2}\neq 0$ and the normalization factor is
\begin{equation}
N=\sqrt{|\alpha c|^{2}+|\beta a|^{2}+2\mathrm{Re}(\alpha c\beta^{*}a^{*}\langle\phi|\psi\rangle)}.
\end{equation}
The entire success probability of the quantum adder is given by
\begin{equation}
P=\dfrac{1}{2}[|\alpha c|^{2}+|\beta a|^{2}+2\mathrm{Re}(\alpha c\beta^{*}a^{*}\langle\phi|\psi\rangle)].
\end{equation}

The above process can be similarly extended to the case in which more than two states are superposed. The key ingredient is to entangle the input states by using control-SWAP operations. For instance, in the case of three states $|\psi\rangle_{1}$, $|\phi\rangle_{2}$, $|\xi\rangle_{3}$, entangled state
\begin{equation}
|\Phi\rangle_{123}\propto |\psi\rangle_{1}|\phi\rangle_{2}|\xi\rangle_{3}+|\phi\rangle_{1}|\psi\rangle_{2}|\xi\rangle_{3}+|\xi\rangle_{1}|\phi\rangle_{2}|\psi\rangle_{3}
\end{equation}
should be firstly generated by states swap between photons $1, 2$ and $1, 3$ respectively. The expected superposed state is then obtained by projecting photon $2$ and $3$ onto some referential states $|\mu\rangle, |\xi\rangle$ that have nonzero overlaps with input states.

\begin{table*}[tbp]
\centering
\caption{Measurement results for 11 pairs of input states. The fidelity $F\equiv (\mathrm{Tr}\sqrt{\rho^{1/2}\sigma\rho^{1/2}})^{2}$ and state distance $D\equiv\dfrac{1}{2}\mathrm{Tr}|\rho-\sigma|$ are calculated as quality measure of the quantum adder.}
\begin{tabular*}{13cm}{@{\extracolsep{\fill}}|c|c|c|c|}
\hline
input state 1  &input state 2  &fidelity  &distance \\   \hline
$|H\rangle$  &$(|H\rangle+|V\rangle)/\sqrt{2}$  &$0.9990\pm 0.0064$  &$0.0311\pm 0.0089$  \\  \hline
$|H\rangle$  &$(\sqrt{2}|H\rangle+|V\rangle)/\sqrt{3}$  &$0.9969\pm 0.0061$ &$0.0552\pm 0.0064$  \\ \hline
$|H\rangle$  &$(|H\rangle+\sqrt{2}|V\rangle)/\sqrt{3}$  &$0.9979\pm 0.0058$ &$0.0460\pm 0.0074$  \\ \hline
$(|H\rangle+|V\rangle)/\sqrt{2}$  &$(\sqrt{2}|H\rangle+|V\rangle)/\sqrt{3}$ &$0.9964\pm 0.0030$ &$0.0594\pm 0.0062$ \\  \hline
$(\sqrt{2}|H\rangle+|V\rangle)/\sqrt{3}$  &$(|H\rangle+\sqrt{2}|V\rangle)/\sqrt{3}$  &$0.9935\pm 0.0043$  &$0.0521\pm 0.0072$  \\ \hline
$(|H\rangle+|V\rangle)/\sqrt{2}$  &$(|H\rangle-|V\rangle)/\sqrt{2}$ &$0.9845\pm 0.0076$  &$0.0226\pm 0.0070$  \\ \hline

$(|H\rangle+|V\rangle)/\sqrt{2}$  &$(|H\rangle+e^{i0.775\pi}|V\rangle)/\sqrt{2}$  &$0.9969\pm 0.0050$  &$0.0552\pm 0.0044$  \\  \hline
$(|H\rangle+|V\rangle)/\sqrt{2}$  &$(|H\rangle+e^{i0.633\pi}|V\rangle)/\sqrt{2}$ &$0.9900\pm 0.0056$  &$0.0335\pm 0.0061$  \\ \hline
$(|H\rangle+|V\rangle)/\sqrt{2}$  &$(|H\rangle+e^{i0.45\pi}|V\rangle)/\sqrt{2}$ &$0.9888\pm 0.0047$  &$0.0475\pm 0.0056$  \\ \hline
$(|H\rangle+|V\rangle)/\sqrt{2}$  &$(|H\rangle+e^{i0.33\pi}|V\rangle)/\sqrt{2}$ &$0.9860\pm 0.0041$  &$0.0434\pm 0.0051$  \\ \hline
$(|H\rangle+|V\rangle)/\sqrt{2}$  &$(|H\rangle+e^{i0.147\pi}|V\rangle)/\sqrt{2}$ &$0.9896\pm 0.0038$  &$0.0455\pm 0.0014$  \\ \hline

\end{tabular*}
\end{table*}

{\it Experimental implementation of the quantum adder.}
Our cross-like quantum adder, which is capable of superposing arbitrary two unknown photonic qubits provided that they have nonzero overlaps with the horizontal polarization state $|H\rangle$, is shown in Fig. 2. It consists of four main stages, i.e., preparation of two unknown qubits, realization of control-SWAP, state projection and coincidence detection.

The polarization state of a pair of photons in state $|\varphi\rangle=|H\rangle\otimes|H\rangle$ is firstly generated by pumping a type-I cut $\beta$-BBO crystal using laser (Topmode 404) with the wavelength at $404$ nm. The pair of photons are then sent into quantum adder from different ports. Arbitrary two qubits $|\psi\rangle_{1}=a|H\rangle+b|V\rangle$ and $|\phi\rangle_{2}=c|H\rangle+d|V\rangle$ can be prepared through combination of half wave plates (HWP1, HWP2) and quarter wave plates (QWP1, QWP2) before beam displacers (BD1, BD2) that separate horizontal and vertical polarization states of photon.

To implement the control-SWAP of states $|\psi\rangle_{1}$ and $|\phi\rangle_{2}$, which is the core component for realizing quantum superposition, we transform polarization qubits to path qubits and use the polarization degree of freedom as control.
The polarization qubits are recovered after the control-SWAP operation. The polarizing beam splitter (PBS1), which transmits $|H\rangle$ and reflects $|V\rangle$, combined with the post-selection of coincident events is used to realize the control-SWAP operation. Polarization qubits are transformed to path qubits by using BD1 and BD2 and setting HWP3 and HWP4 at $22.5^{\circ}$. Before entering PBS1, the state of two photons is
\begin{equation}
|\Phi\rangle_{12}=(a|1\rangle+b|2\rangle)_{1}|+\rangle_{1}\otimes(c|3\rangle+d|4\rangle)_{2}|+\rangle_{2},
\end{equation}
where $|1\rangle, |2\rangle, |3\rangle$, and $|4\rangle$ represent path states and $|\pm\rangle=(|H\rangle\pm|V\rangle)/\sqrt{2}$.
Given that only the coincident events at the final stage of detection is retained, the composite state of the two photons after PBS1 is given by
\begin{equation}
\begin{split}
|\tilde{\Phi}\rangle_{12}=\dfrac{1}{\sqrt{2}}\lbrace&[(a|1\rangle+b|2\rangle)\otimes(c|3\rangle+d|4\rangle)]|H\rangle_{1}|H\rangle_{2}
\\
+&[(a|3\rangle+b|4\rangle)\otimes(c|1\rangle+d|2\rangle)]|V\rangle_{1}|V\rangle_{2}\rbrace,
\end{split}
\end{equation}
with success probability of $1/2$. The control-SWAP operation in our case is not unitary since post-selection is involved and the control qubit used here is $(|H\rangle_{1}|H\rangle_{2}+|V\rangle_{1}|V\rangle_{2})/\sqrt{2}$. The path states of photons remain unchanged when they are in polarization state $|H\rangle|H\rangle$, while their path states are swapped in polarization state $|V\rangle|V\rangle$.

The control degree of freedom is then erased by setting HWP5 and HWP6 at $\theta/2$ and only choosing state $|H\rangle$ with PBS2 and PBS3. The path qubits can be transformed back to polarization qubits after passing through BD3 and BD4 with HWP7 and HWP9 fixed at $45^{\circ}$ \cite{Kwiat}. The composite state of the two photons now becomes (unnormalized)
\begin{equation}
|\Upsilon\rangle_{12}=(\mathrm{cos}^{2}\theta|\psi\rangle_{1}|\phi\rangle_{2}+\mathrm{sin}^{2}\theta|\phi\rangle_{1}|\psi\rangle_{2})|5\rangle_{1}|6\rangle_{2}.
\end{equation}
Projecting photon 2 onto referential state $|H\rangle$, we can obtain the expected state of superposition
\begin{equation}
|\Psi\rangle_{1}=\dfrac{1}{N_{\Psi}}(c\mathrm{cos}^{2}\theta|\psi\rangle+a\mathrm{sin}^{2}\theta|\phi\rangle)
\end{equation}
on photon 1, where
\begin{equation}
 N_{\Psi}=\sqrt{(c\mathrm{cos}^{2}\theta)^{2}+(a\mathrm{sin}^{2}\theta)^{2}+\dfrac{1}{2}\mathrm{sin}^{2}(2\theta)\mathrm{Re}(a^{*}c\langle\phi|\psi\rangle)}.
 \end{equation}

The detection of coincidence implies the success of superposing two unknown qubits and the total success probability can be calculated as
\begin{equation}
P_{\Psi}=\dfrac{1}{2}N^{2}_{\Psi}.
\end{equation}
The entire success probability depends on the success probability of realizing control-SWAP operation, projection probability and input states, which can not be specified for unknown input states. Although the success probability of the quantum adder is not theoretically optimal, it does not matter in our situation because of the quantum adder is heralded by coincident events.

{\it Testing the quantum adder.}
We input some certain qubit pairs to test the quality of quantum adder. The output state $\rho$ is reconstructed by tomography \cite{tom} and compared to the theoretical expected state $|\Psi\rangle$ of superposition by calculating the fidelity $F\equiv (\mathrm{Tr}\sqrt{\rho^{1/2}\sigma\rho^{1/2}})^{2}$ and the state distance $D\equiv\dfrac{1}{2}\mathrm{Tr}|\rho-\sigma|$ with $\sigma=|\Psi\rangle\langle\Psi|$ \cite{Chuang}.

The parameter $\theta$ in our test is chosen to be $45^{\circ}$ so that the theoretical expected state of superposition is
\begin{equation}
|\Psi\rangle=\dfrac{1}{N}(c|\psi\rangle+a|\phi\rangle),
\end{equation}
where $N=\sqrt{|a|^{2}+|c|^{2}+2\mathrm{Re}(a^{*}c\langle\phi|\psi\rangle)}$. Two different classes of qubit pairs, including qubit pairs from the equator of Bloch sphere and qubit pairs without relative phase, are selected (Table I). In the case of qubit pairs chosen from the equator of the Bloch sphere, we fix one input qubit in state $(|H\rangle+|V\rangle)/\sqrt{2}$ and vary the other qubit state along the equator. Note that two input qubits with the same state is trivial in the test.

As shown in Fig. 2, the two photons are sent into a Hong-Ou-Mandel interferometer with visibility of $0.996\pm0.001$ \cite{liu}. This interferometer consists of four BD version Mach-Zehnder interferometers (MZIs). For each MZI, we observe an interference fringe with the visibility above 0.98. The visibility of the post-selection state $(|H\rangle_{1}|H\rangle_{2}+|V\rangle_{1}|V\rangle_{2})/\sqrt{2}$ after passing through PBS1 is measured to be $0.984\pm0.001$ in the basis of $\lbrace(|H\rangle\pm|V\rangle)/\sqrt{2}\rbrace$. For each run, we record clicks for $60$ s and the total coincidence counts are about $5000$.

We test total 11 different input qubit pairs and the measurement results are shown in Table I. The fidelity and distance between measured output state and theoretical expected state of superposition, which are adopted as quality measure of the quantum adder, are calculated according to tomography results. The average fidelity is as high as 0.99 and the average distance is 0.04, which shows the excellent reliability of our heralded quantum adder. It is worth to note that the realization of the key operation, i.e., control-SWAP (Fredkin gate), is different from other schemes, e.g., in Ref. \cite{Fred1}. We use two polarization qubits $\alpha|H\rangle|H\rangle+\beta|V\rangle|V\rangle$ in our scheme as the single control qubit $\alpha|0\rangle+\beta|1\rangle$, which promotes the success probability of the linear optical Fredkin gate to $1/2$.

{\it Summary.}
In conclusion, we have reported a heralded probabilistic quantum adder capable of superposing arbitrary two unknown qubits given that they have nonzero overlaps with the referential state $|H\rangle$. The average fidelity of 0.99 in the test shows the high reliability of our heralded quantum adder. The success probability of practical quantum adder deeply relies on the performance of the control-SWAP operation which is still a great challenge till now \cite{Fred1, Fred2, Fred3}. By coupling light to atoms in cavity \cite{cavity1, cavity2}, trapped ions \cite{trap} etc, the quantum adder can also realize superposition of qubits encoded in other physical systems. The realization of quantum adder not only can deepen our understanding of the fascinating concept of quantum superposition, but also shows great potential applications in quantum information and quantum computation, such as preparing non-classical states and realizing information compression by coherent superposing results of independent runs.

{\it Acknowledgements.}
This project was supported by the National Natural Science Foundation of China (Grant Nos. 61275122, 61590932, 61327901, 11374288, 11474268, 61225025, 11274289 and 11325419), the Strategic Priority Research Program (B) of the Chinese Academy of Sciences (Grant Nos. XDB01030200 and XDB01030300), the Fundamental Research Funds for the Central Universities of China (Grant Nos. WK2470000018 and WK2470000022), the National High-level Talents Support Program (No. BB2470000005).

\end{document}